# ECONOMETRICS FOR DECISION MAKING:
## Building Foundations Sketched by Haavelmo and Wald

Charles F. Manski
Department of Economics and Institute for Policy Research
Northwestern University

Abstract

Haavelmo (1944) proposed a probabilistic structure for econometric modeling, aiming to make econometrics useful for decision making. His fundamental contribution has become thoroughly embedded in subsequent econometric research, yet it could not answer all the deep issues that the author raised. Notably, Haavelmo struggled to formalize the implications for decision making of the fact that models can at most approximate actuality. In the same period, Wald (1939, 1945) initiated his own seminal development of statistical decision theory. Haavelmo favorably cited Wald, but econometrics did not embrace statistical decision theory. Instead, it focused on study of identification, estimation, and statistical inference. This paper proposes statistical decision theory as a framework for evaluation of the performance of models in decision making. I particularly consider the common practice of *as-if optimization*: specification of a model, point estimation of its parameters, and use of the point estimate to make a decision that would be optimal if the estimate were accurate. A central theme is that one should evaluate as-if optimization or any other model-based decision rule by its performance across the state space, listing all states of nature that one believes feasible, not across the model space. I apply the theme to prediction and treatment choice. Statistical decision theory is conceptually simple, but application is often challenging. Advancement of computation is the primary task to continue building the foundations sketched by Haavelmo and Wald.

An early draft of this paper provided the source material for my Haavelmo Lecture at the University of Oslo, December 3, 2019. The paper supersedes one circulated in January and September 2019 under the draft title "Statistical Inference for Statistical Decisions." I am grateful to Valentyn Litvin for excellent research assistance and comments. I am grateful to Olav Bjerkholt, Ivan Canay, Gary Chamberlain, Michael Gmeiner, Kei Hirano, Joel Horowitz, Guido Imbens, Matt Masten, Bruce Spencer, Alex Tetenov, and anonymous reviewers for helpful comments.

## 1. Introduction: Joining Haavelmo and Wald

Early in the development of econometrics, Trygve Haavelmo compared astronomy and planning to differentiate two objectives for modeling: to advance science and to inform decision making. He wrote (Haavelmo, 1943, p. 10):

> "The economist may have two different purposes in mind when he constructs a model . . . . First, he may consider himself in the same position as an astronomer; he cannot interfere with the actual course of events. So he sets up the system . . . . as a tentative description of the economy. If he finds that it fits the past, he hopes that it will fit the future. On that basis he wants to make predictions, assuming that no one will interfere with the game. Next, he may consider himself as having the power to change certain aspects of the economy in the future. If then the system . . . . has worked in the past, he may be interested in knowing it as an aid in judging the effect of his intended future planning, because he thinks that certain elements of the old system will remain invariant."

Jacob Marschak, supporting Haavelmo's work, made a related distinction between meteorological and engineering types of inference; see Bjerkholt (2010) and Marschak and Andrews (1944).

Comparing astronomy and planning provides a nice metaphor for two branches of econometrics. In 1943, before space flight, an astronomer might model a solar system to advance physical science, but the effort could have no practical impact on decision making. An economist might similarly model an economy to advance social science. However, an economist might also model to inform society about the consequences of contemplated decisions that would change aspects of the economy.

Haavelmo's doctoral thesis (Haavelmo, 1944) proposed a probabilistic structure for econometrics that aimed to make it useful for public decision making. To conclude, he wrote (p. 114-115):

> "In other quantitative sciences the discovery of "laws," even in highly specialized fields, has moved from the private study into huge scientific laboratories where scores of experts are engaged, not only in carrying out actual measurements, but also in working out, with painstaking precision, the formulae to be tested and the plans for crucial experiments to be made. Should we expect less in economic research, if its results are to be the basis for economic policy upon which might depend billions of dollars of national income and the general economic welfare of millions of people?"

Haavelmo's thesis made fundamental contributions that became thoroughly embedded in econometrics. Nevertheless, it is unsurprising that it did not answer all the deep issues that the author raised. Notably, Haavelmo struggled to formalize the implications for decision making of the fact that models only approximate actuality. He called attention to this in his opening chapters on "Abstract Models and Reality" and "The Degree of Permanence of Economic Laws," but the later chapters did not resolve the matter.

Haavelmo devoted a chapter to "The Testing of Hypotheses," expositing the then recent work of Neyman and Pearson (1928, 1933) and considering its potential use to evaluate the consistency of models with observed data. Testing models subsequently became widespread, both as a topic of study in econometric theory and as a practice in empirical research. However, Neyman-Pearson testing does not provide satisfactory guidance for decision making. See Section 2.3 below.

While Haavelmo was writing his thesis, Abraham Wald was initiating his own seminal development of statistical decision theory in Wald (1939, 1945) and elsewhere, which culminated in his treatise (Wald, 1950). Wald's work has broad potential application. Indeed, it implicitly provides a framework for evaluation of the use of models in decision making. I say that Wald "implicitly" provides this framework because, writing abstractly, he appears not to have explicitly examined decision making with models. Yet it is straightforward to use statistical decision theory this way. Explaining this motivates the present paper.

I find it intriguing to join the contributions of Haavelmo and Wald because they interacted in the United States during the wartime period when both were developing their ideas. Wald came to the U.S. in 1938 as a refugee from Austria. Haavelmo arrived in 1939 for what was intended to be a short visit, but which lasted the entire war when he could not return to occupied Norway. Bjerkholt (2007, 2015) describes the many interactions of Haavelmo and Wald, not only at conferences but also in hiking expeditions.

Haavelmo's appreciation of Wald is clear. In the preface of Haavelmo (1944), he wrote (p. v):

> "My most sincere thanks are due to Professor Abraham Wald of Columbia University for numerous suggestions and for help on many points in preparing the manuscript. Upon his unique knowledge of modern statistical theory and mathematics in general I have drawn very heavily. Many of the statistical sections in this study have been formulated, and others have been reformulated, after discussions with him."

The text of the thesis cites several of Wald's papers. Most relevant is the final chapter on "Problems of Prediction," where Haavelmo suggests application of the framework in Wald (1939) to choose a predictor of a random outcome. I discuss this in Section 4.1 below.

Despite Haavelmo's favorable citations of Wald's ideas, econometrics did not embrace statistical decision theory. Instead, it focused on identification, estimation, and statistical inference. No contribution in Cowles Monograph 10 (Koopmans, 1950) mentions statistical decision theory. Only one does so briefly in Cowles Monograph 14 (Hood and Koopmans, 1953). This appears in a chapter by Koopmans and Hood (1953), who refer to estimates of structural parameters as "raw materials, to be processed further into solutions of a wide variety of prediction problems." See Section 4.1 for further discussion.

Modern econometricians continue to view parameter estimates as "raw materials" that may be used to solve decision problems. A widespread practice has been *as-if* optimization: specification of a model, point estimation of its parameters, and use of the point estimate to make a decision that would be optimal if the estimate were accurate. As-if optimization, also called *plug-in* or *two-step* decision making, has heuristic appeal when a model is known to be correct but less so when the model may be incorrect.

A huge hole in econometric theory has been the absence of a broadly applicable means to evaluate as-if opti mization and other uses of econometric models in decision making. This paper proposes statistical decision theory as a framework for evaluating the performance of models. I set forth the general idea and apply it to two prevalent decision problems, prediction and treatment choice.

Section 2 reviews basic elements of statistical decision theory and introduces notation used throughout the paper. Section 3 shows how the Wald framework may be used to evaluate decision making with models. One specifies a model space, which approximates the state space in some way. A model-based decision uses the model space as if it were the state space. I consider the use of models to perform as-if optimization. My theme is that one should evaluate as-if optimization or any other model-based decision rule by its performance across the state space, not the model space. Thus, statistical decision theory embraces use of both correct and incorrect models to make decisions. I relate this idea to research on estimation of misspecified models, specification tests, and robust decisions, explaining the connections and differences.

Sections 4 and 5 consider two decision problems that have long been central to econometrics, prediction and treatment choice. Both subjects have drawn substantial attention from the conditional Bayes perspective, but not nearly enough using the Wald framework. A small body of work has evaluated the maximum regret of decision criteria. I present new computational findings on prediction and new analytical and computational findings on treatment choice.

Considered broadly, this paper adds to the argument made beginning in Manski (2000) and then in a sequence of subsequent articles for application of Wald's statistical decision theory to econometrics. A group of econometricians have made recent contributions towards this objective. The main focus of this work has been treatment choice with data from randomized trials, with contributions by Manski (2004, 2005, 2019), Hirano and Porter (2009, 2019), Stoye (2009, 2012), Manski and Tetenov (2016, 2019, 2021), and Kitagawa and Tetenov (2018). Manski (2007) and Athey and Wager (2019) have studied treatment choice with observational data. Chamberlain (2000a, 2007) and Chamberlain and Moreira (2009) have used statistical decision theory to study estimation of some linear econometric models. Dominitz and Manski (2017, 2021) have studied prediction with missing data.

The original contributions made here are varied. New perspective on the history of econometric thought permeates the paper. The idea proposed in Section 3---use of the Wald framework to evaluate as-if optimization and other model-based decision rules, measuring performance across the state space rather than model space---is obvious in retrospect. Yet it has not been widely appreciated. The paper reports new analyses of prediction and treatment choice as statistical decision problems in Sections 4 and 5.

Before we begin, I think it important to emphasize that Wald's statistical decision theory is prescriptive rather than descriptive. Prescriptive decision analysis seeks to develop reasonable, even if not optimal, ways to make decisions under uncertainty. Descriptive analysis seeks to understand and predict how actual decision makers behave. See Bell, Raiffa, and Tversky (1988) for extensive discussion of this important distinction. This paper does not perform or discuss descriptive research on decision making.

## 2. Basic Elements of Statistical Decision Theory

The Wald development of statistical decision theory directly addresses decision making with sample data. Wald began with the standard problem of a planner (or decision maker or agent) who must choose an action yielding welfare that depends on an unknown state of nature. The planner specifies a state space listing the states considered possible. He chooses without knowing the true state.

Wald added to this standard problem by supposing that the planner observes sample data that may be informative about the true state. He studied choice of *a statistical decision function (SDF),* which maps each potential data realization into a feasible action. He proposed evaluation of SDFs as procedures, chosen ex ante, specifying how a planner would use whatever data are realized. Thus, Wald's theory is frequentist.

### 2.1. Decisions Under Uncertainty

I describe decision problems without sample data here and with such data next. Consider a planner who must choose an action yielding welfare that varies with the state of nature. The planner has an objective function and beliefs about the feasible values of the true state. These are considered primitives. He must choose an action without knowing the true state.

Formally, the planner faces choice set C and believes that the true state lies in set S, called the state space. The objective function $w(\cdot, \cdot): C \times S \to R^1$ maps actions and states into welfare. The planner ideally would maximize $w(\cdot, s^*)$, where $s^*$ is the true state. However, he only knows that $s^* \in S$.

The choice set is commonly considered to be predetermined. The welfare function and the state space are subjective. The former formalizes what the planner wants to achieve, and the latter expresses the states he believes could possibly occur. While the state space ultimately is subjective, its structure may use data that are informative, eliminating some states as possibilities. This idea is central to analysis of identification.

Haavelmo considered the state space to be a set of probability distributions that may possibly describe the system under study. The Koopmans (1949) formalization of identification contemplated unlimited data collection that enables one to shrink the state space, eliminating distributions that are inconsistent with the information revealed by observation. The true state is point identified if maintained assumptions and the observational process eliminate all but one distribution for the economic system. It is partially identified if assumptions and observation eliminate some but not all distributions initially deemed possible.

Given a welfare function and state space, a close to universally accepted prescription for decision making is that choice should respect dominance. Action $c \in C$ is weakly dominated if there exists a $d \in C$ such that $w(d, s) \geq w(c, s)$ for all $s \in S$ and $w(d, s) > w(c, s)$ for some $s \in S$. Even though the true state $s^*$ is unknown, choice of d is certain to weakly improve on choice of c.

There is no clearly best way to choose among undominated actions, but decision theorists have not wanted to abandon the idea of optimization. So they have proposed various ways of using the objective function $w(\cdot, \cdot)$ to form functions of actions alone, which can be optimized. In principle one should only consider undominated actions, but it often is difficult to determine which actions are undominated. Hence, in practice it is common to optimize over the full set of feasible actions. I define decision criteria accordingly in this paper. I also use max and min notation, without concern for the mathematical subtleties that sometimes make it necessary to use sup and inf operations.

One idea is to place a subjective probability distribution on the state space, average state-dependent welfare with respect to this distribution, and maximize the resulting function. This yields maximization of subjective average welfare. Let $\pi$ be the specified distribution on S. For each feasible action c, $\int w(c, s)d\pi$ is the mean of $w(c, s)$ with respect to $\pi$. The criterion solves the problem

$$(1) \quad \max_{c \in C} \int w(c, s)d\pi.$$

Another idea is to seek an action that, in some sense, works uniformly well over all elements of S. This yields the maximin and minimax-regret (MMR) criteria. The maximin criterion maximizes the minimum welfare attainable across the elements of S. For each feasible action c, consider the minimum feasible value of $w(c, s)$; that is, $\min_{s \in S} w(c, s)$. A maximin rule chooses an action that solves

$$\max_{c \in C} \min_{s \in S} w(c, s). \tag{2}$$

The MMR criterion chooses an action that minimizes the maximum loss to welfare that can result from not knowing the true state. An MMR choice solves

$$\min_{c \in C} \max_{s \in S} [\max_{d \in C} w(d, s) - w(c, s)]. \tag{3}$$

Here $\max_{d \in C} w(d, s) - w(c, s)$ is the *regret* of action c in state s; that is, the welfare loss associated with choice of c relative to an action that maximizes welfare in state s. The true state being unknown, one evaluates c by its maximum regret over all states and selects an action that minimizes maximum regret. The maximum regret of an action measures its maximum distance from optimality across states. Hence, an MMR choice is uniformly nearest to optimal among the feasible actions.

A planner who asserts a partial subjective distribution on the state space, placing lower and upper probabilities on states as in Dempster (1968) or Walley (1991), could maximize minimum subjective average welfare or minimize maximum average regret. These hybrid criteria combine elements of averaging across states and concern with uniform performance across states. Statistical decision theorists refer to these criteria as Γ-maximin and Γ-minimax regret (Berger, 1985). The former has drawn attention from axiomatic decision theorists, calling it maxmin expected utility (Gilboa and Schmeidler, 1989). See Chamberlain (2000b) for an application in econometrics. I will confine discussion to the polar cases in which the planner asserts a complete subjective distribution or none.

### 2.2. Statistical Decision Problems

Statistical decision problems add to the above structure by supposing that the planner observes data generated by some sampling distribution. Sample data may be informative but, unlike the unlimited data contemplated in identification analysis, they do not enable one to shrink the state space.

Knowledge of the sampling distribution is generally incomplete. To express this, one extends the concept of the state space S to list the set of feasible sampling distributions, denoted $(Q_s, s \in S)$. Let $\Psi_s$ denote the sample space in state s; that is, $\Psi_s$ is the set of samples that may be drawn under sampling distribution $Q_s$. The literature typically assumes that the sample space does not vary with s and is known. I maintain this assumption and denote the sample space as $\Psi$, without the s subscript. Then a statistical decision function $c(\cdot)$: $\Psi \rightarrow C$ maps the sample data into a chosen action.

Wald's concept of a statistical decision function embraces all mappings [data → action]. An SDF need not perform inference; that is, it need not use data to draw conclusions about the true state of nature. The prominent decision criteria that have been studied — maximin, minimax-regret, and maximization of subjective average welfare — do not refer to inference. The general absence of inference in statistical decision theory is striking and has been noticed; see Neyman (1962) and Blyth (1970).

Although SDFs need not perform inference, some do. That is, some have the sequential form [data → inference → action], first performing inference and then using the inference to make a decision. There seems to be no accepted term for such SDFs, so I call them *inference-based*.

SDF $c(\cdot)$ is a deterministic function after realization of the sample data, but it is a random function ex ante. Hence, the welfare achieved by $c(\cdot)$ is a random variable ex ante. Wald's theory evaluates the performance of $c(\cdot)$ in state s by $Q_s\{w[c(\psi), s]\}$, the ex-ante distribution of welfare that it yields across realizations $\psi$ of the sampling process.

It remains to ask how planners might compare the welfare distributions yielded by different SDFs. Planners want to maximize welfare, so it seems self-evident that they should prefer SDF $d(\cdot)$ to $c(\cdot)$ in state s if $Q_s\{w[d(\psi), s]\}$ stochastically dominates $Q_s\{w[c(\psi), s]\}$. It is less obvious how they should compare SDFs whose welfare distributions do not stochastically dominate one another.

Wald proposed measurement of the performance of $c(\cdot)$ in state s by its expected welfare across samples; that is, $E_s\{w[c(\psi), s]\} \equiv \int w[c(\psi), s]dQ_s$. An alternative that has drawn only slight attention measures performance by quantile welfare (Manski and Tetenov, 2014). Writing in a context where one wants to minimize loss rather than maximize welfare, Wald used the term *risk* to denote the mean performance of an SDF across samples.

In practice, one does not know the true state. Hence, one evaluates $c(\cdot)$ by the state-dependent expected welfare vector $(E_s\{w[c(\psi), s]\}, s \in S)$. Using the term *inadmissible* to denote weak dominance when evaluating performance by risk, Wald recommended elimination of inadmissible SDFs from consideration. As in decisions without sample data, there is no clearly best way to choose among admissible SDFs.

Statistical decision theory has mainly studied the same decision criteria as has decision theory without sample data. Let $\Gamma$ be a specified set of feasible SDFs, each mapping $\Psi \rightarrow C$. The statistical versions of decision criteria (1), (2), and (3) are

(4) $$\max_{c(\cdot) \in \Gamma} \int E_s\{w[c(\psi), s]\}\, d\pi,$$

(5) $$\max_{c(\cdot) \in \Gamma} \min_{s \in S} E_s\{w[c(\psi), s]\},$$

(6) $$\min_{c(\cdot) \in \Gamma} \max_{s \in S} \left( \max_{d \in C} w(d, s) - E_s\{w[c(\psi), s]\}\right).$$

An Appendix discusses these criteria further.

## 2.3. Binary Choice Problems

SDFs for binary choice problems are simple and interesting. They can always be viewed as hypothesis tests. Yet the Wald perspective on testing differs considerably from that of Neyman-Pearson.

Let choice set C contain two actions, say {a, b}. A SDF $c(\cdot)$ partitions $\Psi$ into two regions that separate the data yielding choice of each action. These are $\Psi_{c(\cdot)a} \equiv [\psi \in \Psi\colon c(\psi) = a]$ and $\Psi_{c(\cdot)b} \equiv [\psi \in \Psi\colon c(\psi) = b]$.

A test motivated by the choice problem partitions state space S into two regions, say $S_a$ and $S_b$, that separate the states in which actions a and b are uniquely optimal. Thus, $S_a$ contains the states [s ∈ S: w(a, s) > w(b, s)] and $S_b$ contains [s ∈ S: w(b, s) > w(a, s)]. The choice problem does not provide a rationale for allocation of states in which the actions yield equal welfare. The standard practice is to give one action, say a, a privileged status and to place all states yielding equal welfare in $S_a$. Then $S_a \equiv [s \in S\text{: } w(a, s) \geq w(b, s)]$ and $S_b \equiv [s \in S\text{: } w(b, s) > w(a, s)]$.

In the language of testing, SDF c(·) performs a test with acceptance regions $\Psi_{c(\cdot)a}$ and $\Psi_{c(\cdot)b}$. When $\psi \in \Psi_{c(\cdot)a}$, c(·) accepts the hypothesis {s ∈ $S_a$} by setting c(ψ) = a. When $\psi \in \Psi_{c(\cdot)b}$, c(·) accepts the hypothesis {s ∈ $S_b$} by setting c(ψ) = b. I use the word "accepts" rather than the traditional term "does not reject" because choice of a or b is an affirmative action.

Although all SDFs for binary choice are interpretable as tests, Neyman-Pearson testing and statistical decision theory evaluate tests in different ways, explained below.

### 2.3.1. Neyman-Pearson Testing

Neyman and Pearson (1928, 1933) tests view the states {s ∈ $S_a$} and {s ∈ $S_b$} asymmetrically, calling the former the null hypothesis and the latter the alternative. The sampling probability of rejecting the null when it is correct is the probability of a Type I error. A longstanding convention has been to restrict attention to tests in which the probability of a Type I error is no larger than a predetermined value α, usually 0.05, for all s ∈ $S_a$. Thus, one restricts attentions to SDFs c(·) for which $Q_s[c(\psi) = b] \leq \alpha$ for all s ∈ $S_a$.

Among tests that satisfy this restriction, it is thought desirable to use ones that have small probability of rejecting the alternative hypothesis when it is correct, called the probability of a Type II error. However, it generally is not possible to attain small probability of a Type II error for all s ∈ $S_b$. Letting S be a metric space, the probability of Type II error typically approaches 1 − α as s ∈ $S_b$ nears the boundary of $S_a$. The practice has been to restrict attention to states in $S_b$ that lie at least a specified distance from $S_a$. Let ρ be the metric on S. Let $\rho_a > 0$ be the specified minimum distance from $S_a$. Neyman-Pearson testing seeks small values for the maximum value of $Q_s[c(\psi) = a]$ over $s \in S_b$ s. t. $\rho(s, S_a) \geq \rho_a$.

### 2.3.2. Expected Welfare of Tests

Decision theoretic evaluation of tests does not restrict attention to tests that yield a predetermined upper bound on the probability of a Type I error. Nor does it aim to minimize the maximum value of the probability of a Type II error when more than a specified minimum distance from the null hypothesis. Wald proposed for binary choice, as elsewhere, evaluation of the performance of SDF $c(\cdot)$ in state s by the expected welfare that it yields across realizations of the sampling process. He first addressed testing this way in Wald (1939).

The welfare distribution in state s in a binary choice problem is Bernoulli, with mass points max [w(a, s), w(b, s)] and min [w(a, s), w(b, s)]. These coincide if w(a, s) = w(b, s). When w(a, s) ≠ w(b, s), let $R_{c(\cdot)s}$ denote the probability that $c(\cdot)$ yields an error, choosing the inferior action over the superior one. That is,

$$
\begin{aligned}
(7) \qquad R_{c(\cdot)s} &= Q_s[c(\psi) = b] \quad \text{if } w(a, s) > w(b, s), \\
&= Q_s[c(\psi) = a] \quad \text{if } w(b, s) > w(a, s).
\end{aligned}
$$

The former and latter are the probabilities of Type I and Type II errors.

The probabilities that welfare equals max [w(a, s), w(b, s)] and min [w(a, s), w(b, s)] are $1 - R_{c(\cdot)s}$ and $R_{c(\cdot)s}$. Hence, expected welfare in state s is

$$
\begin{aligned}
(8) \qquad E_s\{w[c(\psi), s]\} &= R_{c(\cdot)s}\{\min [w(a, s), w(b, s)]\} + [1 - R_{c(\cdot)s}]\{\max [w(a, s), w(b, s)]\} \\
&= \max [w(a, s), w(b, s)] - R_{c(\cdot)s}\cdot|w(a, s) - w(b, s)|.
\end{aligned}
$$

The expression $R_{c(\cdot)s}\cdot|w(a, s) - w(b, s)|$ is the regret of $c(\cdot)$ in state s. Thus, regret is the product of the error probability and the magnitude of the welfare loss when an error occurs.

Evaluation of tests by expected welfare constitutes a fundamental difference between the perspectives of Wald and Neyman-Pearson. Planners should care about more than the probabilities of Type I and II error. They should care as well about the magnitudes of the losses to welfare that arise when errors occur. A given error probability should be less acceptable when the welfare difference between actions is larger. Neyman-Pearson theory does not recognize this.

Computation of regret in a specified state is usually tractable. The welfare magnitudes w(a, s) and w(b, s) are usually easy to compute. The error probability $R_{c(\cdot)s}$ may not have an explicit form, but it can be approximated to any desired precision by Monte Carlo integration. One draws repeated pseudo-realizations of $\psi$ from distribution $Q_s$, computes the fraction of cases in which the resulting $c(\psi)$ selects the inferior action, and uses this to estimate $R_{c(\cdot)s}$.

Whereas computation of regret in one state is tractable, computation of maximum regret across all feasible states may be burdensome. The state space commonly is uncountable in applications. A pragmatic process for coping with uncountable state spaces is to discretize the space, computing regret on a finite subset of states that reasonably approximate the full state space.

### 2.3.3. Maximum Regret of Medical Decisions using Neyman-Pearson Tests and the Empirical Success Rule

Medical decision making illustrates well the difference between Neyman-Pearson testing and statistical decision theory. A core objective of randomized trials comparing medical treatments is to inform treatment choice. Often the objective is to compare an existing treatment, called *standard care*, with an innovation. The prevailing statistical practice has been to conclude that the innovation is better than standard care only if a Neyman-Pearson test rejects the null hypothesis that the innovation is no better than standard care.

Manski and Tetenov (2016, 2019, 2020) compare the maximum regret of treatment choice using common Neyman-Pearson tests with decisions using the *empirical success* rule, which chooses a treatment that maximizes the average sample outcome in the trial. The simplest context is choice between two treatments, t = a and t = b, when the outcome of interest is binary, y(t) = 1 denoting success and y(t) = 0 failure. State s indexes a possible value for the pair of outcome probabilities $\{P_s[y(a) = 1], P_s[y(b) = 1]\}$. The welfare yielded by treatment t in state s is $w(t, s) = P_s[y(t) = 1]$. The regret of SDF c(·) in state s is $R_{c(\cdot)s} \cdot |P_s[y(a) = 1] - P_s[y(b) = 1]|$.

We suppose that the planner has no a priori knowledge of the outcome probabilities. Hence, the state space is $[0, 1]^2$. We approximate maximum regret by computing regret over a grid of states, discretizing the state space. Examining a wide range of sample sizes and designs, we find that the empirical success rule yields results that are always much closer to optimal than those generated by common tests.

## 3. Decision Making with Models

### 3.1. Basic Ideas

I stated at the outset that decision theory begins with a planner who specifies a state space listing the states considered possible. Thus, the state space should include all states that the planner believes feasible and no others. The state space may be a large set that is difficult to contemplate in its entirety. Hence, it is common to make decisions using a model.

The word "model" is used informally to connote a simplification or approximation of reality. Formally, a model specifies an alternative to the state space. Thus, model m replaces S with a model space $S_m$. A planner using a model acts as if the model space is the state space. The planner might solve problem (4), (5), or (6) with $S_m$ replacing S. Section 3.2 discusses another approach, as-if optimization.

The states contained in a model space may or may not be elements of the state space. George Box famously wrote (Box, 1979): "All models are wrong, but some are useful." The phrase "all models are wrong" indicates that Box was thinking of models that simplify or approximate reality in a way that one believes could not possibly be correct; then $S_m \cap S = \emptyset$. On the other hand, researchers often use models that they believe could be correct but are not necessarily so; then $S_m \subset S$. Economists have long used models of the second type. They have sought to evaluate such models in various ways.

A common practice when using models to predict macroeconomic outcomes is to compare predictions with observed realizations, judging the usefulness of models by the accuracy of the predictions (e.g., Diebold, 2015; Patton, 2019). Measurement of accuracy requires selection of a loss function, typically square or absolute loss. This connects empirical practice with decision theory. However, model evaluation is performed ex post, with respect to observed outcomes, rather than ex ante as in statistical decision theory. This makes the practice fundamentally different.

A common practice in econometric theory poses an estimator that consistently estimates a well-defined estimand when a specified model is correct and characterizes the estimand to which the estimate converges when the model is incorrect in some sense. For example, Goldberger (1968) observed that the least squares estimate of a linear regression model converges to the *best linear predictor* of y on x if $E(y|x)$ is not a linear function of x. White (1982) observed that the maximum likelihood estimate of a model converges to the parameter that minimizes the Kullback-Leibler information criterion if the model is incorrect. Imbens and Angrist (1994) showed that an instrumental-variables estimate of a model of linear homogeneous treatment response converges to a *local-average treatment effect* if this model is incorrect but a certain monotonicity assumption holds. Historically, research of this type has not been connected to statistical decision theory.

A persistent concern of econometric theory has been to determine when models have implications that may potentially be inconsistent with observable data. These models are called testable, refutable, or over-identified. Working within the Neyman-Pearson paradigm, econometricians have developed *specification tests*, which take the null hypothesis to be that the model is correct and the alternative to be that it is incorrect (e.g., Hausman, 1978; White, 1982). Formally, the null is $\{s^* \in S_m\}$ and the alternative is $\{s^* \notin S_m\}$. However, econometricians have struggled to answer a central question raised by Haavelmo (1944) in his opening chapter on "Abstract Models and Reality" and restated succinctly by Masten and Poirier (2021). The latter write (p. 1): "What should researchers do when their baseline model is refuted?" They discuss the many ways that econometricians have sought to answer the question, and they offer new suggestions.

The literatures cited above have not sought to evaluate the ex-ante performance of models in decision making. Statistical decision theory accomplishes this in a straightforward way. What matters is the SDF, say $c_m(\cdot)$, that one chooses using a model. As with any SDF, one measures the performance of $c_m(\cdot)$ by its vector of state-dependent expected welfares $(E_s\{w[c_m(\psi), s]\}, s \in S)$. The relevant states for evaluation of performance are those in the state space S, not those in the model space $S_m$.

Thus, statistical decision theory operationalizes Box's assertion that some models are useful. One should not make an abstract assertion that a model is or is not useful. Usefulness depends on the decision context. Useful model-based decision rules yield acceptably high state-dependent expected welfare across the state space, relative to what is possible in principle.

3.1.1. Research on Robust Decisions

The remainder of this paper fleshes out the above ideas on evaluation of model-based decisions. Before then, I juxtapose these ideas with those expressed in related research on *robust decisions*. This includes the statistical literature on robust estimation and prediction (e.g., Huber, 1964, 1981; Hampel et. al., 1986), the engineering literature on robust control (e.g., Zhou, Doyle, and Glover, 1996), and econometric work on robust macroeconomic modeling (e.g., Hansen and Sargent, 2001, 2008). The idea that the usefulness of a model depends on the decision context has been appreciated in research on robust decisions. A review article by Watson and Holmes (2016) states (p. 466):

> "Statisticians are taught from an early stage that "essentially, all models are wrong, but some are useful" (Box and Draper, 1987). By "wrong" we will take to mean misspecified and by "useful" we will take to mean helpful for aiding actions (taking decisions), or rather a model is not useful if it does not aid any decision."

Research on robust decisions proceeds in a different manner than does statistical decision theory. Rather than begin with specification of a state space, it begins with specification of a model. Having specified the model, a researcher may be concerned that it is not correct. To recognize this possibility, the researcher enlarges the model space locally, using a specified metric to generate a neighborhood of the model space. He then acts as if the locally enlarged model space is correct. Watson and Holmes write (p.

465): “We then consider formal methods for decision making under model misspecification by quantifying stability of optimal actions to perturbations to the model within a neighbourhood of [the] model space.”

Although research on robust decisions differs procedurally from statistical decision theory, one can subsume the former within the latter if one considers the locally enlarged model space to be the state space. It is unclear how often this perspective characterizes what researchers have in mind. Published articles often do not state explicitly whether the constructed neighborhood of the model space encompasses all states that authors deem sufficiently feasible to warrant formal consideration. The models specified in robust decision analyses often make strong assumptions and the generated neighborhoods often relax these assumptions only modestly.

### 3.2. As-If Optimization

A familiar econometric practice specifies a model space, typically called the parameter space. Sample data are used to select a point in the parameter space, called a point estimate of the parameter. The estimation method is motivated by desirable statistical properties that hold if the true state lies within the model space.

*As-if optimization* chooses an action that optimizes welfare as if the estimate is the true state. Econometricians often use the term “plug-in” or “two-step” rather than “as-if.” I prefer “as-if,” which makes explicit that one acts as if the model is correct. Discussion of as-if optimization has a long history. A prominent case is the Friedman and Savage (1948) discussion of as-if expected utility maximization.

As-if optimization is a type of inference-based SDF. Whereas Wald supposed that a planner both performs research and makes a decision, in practice there commonly is separation between the two. Researchers report inferences and planners use them to make decisions. Thus, planners perform the mapping [inference → decision] rather than the more basic mapping [data → decision]. Having researchers report estimates and planners use them as if they are accurate exemplifies this process.

Formally, a point estimate is a function $s(\cdot)$: $\Psi \to S_m$ that maps data into the model space. As-if optimization means solution of the problem $\max_{c \in C} w[c, s(\psi)]$. When as-if optimization yields multiple solutions, one may use some auxiliary rule to choose among them. The result is an SDF $c[s(\cdot)]$, where

$$(9) \quad c[s(\psi)] \in \underset{c \in C}{\text{argmax}}\ w[c, s(\psi)], \quad \psi \in \Psi.$$

A rationale for solving (9) is that selecting a point estimate and using it to maximize welfare is easier than solving problems (4) to (6). However, computational appeal cannot justify this approach to decision making. To motivate as-if optimization, econometricians often cite limit theorems of asymptotic theory that hold if the model is correct. They hypothesize a sequence of sampling processes indexed by sample size N and a corresponding sequence of point estimates $s_N(\cdot)$: $\Psi_N \to S_m$. They show that the sequence is consistent when specified assumptions hold. That is, $s_N(\psi) \to s^*$ as $N \to \infty$, in probability or almost surely. They may prove further results regarding rate of convergence and the limiting distribution of the estimate.

Asymptotic arguments may be suggestive, but they do not prove that as-if optimization provides a well-performing SDF. Statistical decision theory evaluates as-if optimization in state s by the expected welfare $E_s\{w\{c[s(\psi)], s\}\}$ that it yields across samples of size N, not asymptotically. It calls for study of expected welfare across the state space, not the model space.

### 3.2.1. As-If Optimization with Analog Estimates

Econometric research from Haavelmo onward has focused to a considerable degree on a class of problems that connect the state space and the sampling distribution in a simple way. These are problems in which states are probability distributions and the data are a random sample drawn from the true distribution. In such problems, a natural form of as-if optimization is to act as if the empirical distribution of the data is the true population distribution. Thus, one specifies the model space as the set of all possible empirical distributions and uses the observed empirical distribution as the estimate of the true state.

Goldberger (1968) called this the *analogy principle*. He wrote (p. 4): "The *analogy principle* of estimation . . . . proposes that population parameters be estimated by sample statistics which have the same property in the sample as the parameters do in the population." The empirical distribution consistently estimates the population distribution and has further desirable properties. This suggests decision making using the empirical distribution as if it were the true population distribution.

3.2.2. As-If Decisions with Set Estimates

As-if optimization uses data to compute a point estimate of the true state and chooses an action that optimizes welfare as if this estimate is accurate. An obvious, but rarely applied, extension is to use data to compute a set-valued estimate and act as if the set estimate is accurate. Whereas a point estimate $s(\cdot)$ maps data into an element of $S_m$, a set estimate $S(\cdot)$ maps data into a subset of $S_m$.

For example, $S(\cdot)$ could be a confidence set or an analog estimate of the identification region for a partially identified state. Sections 4.2 and 5.3 apply the latter idea to prediction and treatment choice. Decision making using confidence sets as set estimates is a topic for future research.[1]

Given data $\psi$, one could act as if the state space is $S(\psi)$ rather than the larger set S. Specifically, one could solve these data-dependent versions of problems (1) through (3):

(1') $$\max_{c \in C} \int w(c, s) d\pi(\psi),$$

(2') $$\max_{c \in C} \min_{s \in S(\psi)} w(c, s),$$

(3') $$\min_{c \in C} \max_{s \in S(\psi)} [\max_{d \in C} w(d, s) - w(c, s)].$$

[1] A confidence set with coverage probability $\alpha$ is a set estimate $S(\cdot)$ such that $Q_s[\psi: s \in S(\psi)] \geq \alpha$ , all $s \in S$. Researchers have used confidence sets to quantify imprecision of inference, without reference to decision problems. Nevertheless, they could be used to make as-if decisions. When studying this possibility, I expect that it will be productive to abandon the conventional practice of a priori fixing the coverage probability of a confidence set. This practice mirrors Neyman-Pearson testing, which restricts attention to test statistics that yield at most a specified probability of Type I error.

In the case of (1'), $\pi(\psi)$ is a subjective distribution on the set $S(\psi)$.

These as-if problems are generally easier to solve than problems (4) to (6). The as-if problems fix $\psi$ and select one action c, whereas problems (4) to (6) require one to consider all potential samples and choose a decision function $c(\cdot)$. The as-if problems compute welfare values w(c, s), whereas (4) to (6) compute more complex expected welfare values $E_s\{w[c(\psi), s]\}$.[2]

It is important not to confuse as-if decision making with set estimates, as described here, with sensitivity analysis, which does not apply statistical decision theory. A sensitivity analysis computes multiple point estimates under alternative assumptions. One then performs as-if optimization with each estimate, yielding alternative decisions. When the multiple estimates or decisions coincide, researchers may state that the result is "robust," although this meaning of "robust" differs from that in Section 3.1.1. When sensitivity analysis yields multiple disparate decisions, it does not offer a prescription for decision making.

## 4. Prediction with Sample Data

### 4.1. Haavelmo on Prediction

A familiar case of as-if optimization occurs when states are distributions for a real random variable and the decision is to predict the value of a realization drawn from the true distribution. When welfare is measured by square and absolute loss, the best predictors in each state are well-known to be the population mean and median. When the true distribution is not known but data from a random sample are observed, the analogy principle suggests use of the sample average and median as predictors.

[2] An alternative approach replaces S by $S(\psi)$ in the middle operations of (4) to (6), but it does not replace $E_s\{w[c(\psi), s]\}$ by w(c, s) in the innermost operations. This simplifies (4) to (6) by shrinking the state space over which the middle operations are performed. However, it is more complex than (1′) to (3′) for two reasons. It requires choice of a decision function $c(\cdot)$ rather than a single action c, and it must compute $E_s\{w[c(\psi), s]\}$ rather than w(c, s). Chamberlain (2000a) used asymptotic considerations to suggest this type of as-if decision and presented an application.

In his chapter on "Problems of Prediction," Haavelmo (1944) questioned this common application of as-if optimization and instead recommended application of the Wald theory. In his section on "General Formulation of the Problem of Prediction," he wrote (p. 109): "We see therefore that the seemingly logical 'two-step' procedure of first estimating the unknown distribution of the variables to be predicted and then using this estimate to derive a prediction formula for the variables may not be very efficient." Citing Wald (1939), he proposed computation of the state-dependent risk for any proposed predictor function.

Letting $E_2$ denote a predictor function and $(x_1, x_2, \ldots, x_N)$ be the sample data, he wrote (p. 109): "We have to choose $E_2$ as a function of $x_1, x_2, \ldots, x_N$, and we should, naturally, try to choose $E_2(x_1, x_2, \ldots, x_N)$ in such a way that r (the 'risk') becomes as small as possible." He recognized that there generally does not exist a predictor function that minimizes risk across all states of nature. Hence, he suggested a feasible approach. I quote in full this key passage, which uses the notation $\Omega_1$ to denote the state space (p. 116):

> "In general, however, we may expect that no uniformly best prediction function exists. Then we have to introduce some additional principles in order to choose a prediction function. We may then, first, obviously disregard all those prediction functions that are such that there exists another prediction function that makes r smaller for every member of $\Omega_1$. If this is not the case we call the prediction function considered an admissible prediction function. To choose between several admissible prediction functions we might adopt the following principle, introduced by Wald: For every admissible prediction function $E_2$ the 'risk' r is a function of the true distribution p. Consider that prediction function $E_2$, among the admissible ones, for which the largest value of r is at a minimum (i.e., smaller than or at most equal to the largest value of r for any other admissible $E_2$). Such a prediction function, if it exists, may be said to be the least risky among the admissible prediction functions."

Thus, following Wald, Haavelmo suggested elimination of inadmissible predictors followed by choice of a minimax predictor among those that are admissible.

It may be that econometrics would have progressed to make productive use of statistical decision theory if Haavelmo had been able to pursue this idea further. However, in his next section on "Some Practical Suggestions for the Derivation of Prediction Formulae," he cautioned regarding the practicality of the idea, writing (p. 111): "The apparatus set up in the preceding section, although simple in principle, will in general involve considerable mathematical problems and heavy algebra."

Aiming for tractability, Haavelmo sketched an example of as-if optimization. He noted that one could study the state-dependent risk of the resulting SDF, but he did not provide analysis. With this, his chapter on prediction ended. Thus, Haavelmo initiated econometric consideration of statistical decision theory but, stymied by computational intractability, he found himself unable to follow through.

Nor did other econometricians follow through in the period after publication of Haavelmo (1944). I observed earlier that no contribution in Cowles Monograph 10 mentioned statistical decision theory and only one did so briefly in Cowles 14. Cowles 10 and 14 contain several chapters by Haavelmo and by Wald, but these concern different subjects. The only mention in Cowles 14 appeared in Koopmans and Hood (1953). Considering "The Purpose of Estimation," they wrote (p. 127):

> "if a direct prediction problem . . . . can be isolated and specified, the choice of a method of estimation should be discussed in terms of desired properties of the joint distribution of the prediction(s) made and the realized values(s) of the variables(s) predicted. In particular, in a precisely defined prediction problem of this type, one may know the consequence of various possible errors of prediction and would then be able to use predictors minimizing the mathematical expectation of losses due to such errors. Abraham Wald [1939, 1945, 1950c], among others, has proposed methods of statistical decision-making designed to accomplish this."

However, they went on to state that neither they nor other contributors to Cowles 14 apply statistical decision theory to prediction. They wrote (p. 127):

> "The more classical methods of estimation applied in this volume are not as closely tailored to any one particular prediction problem. Directed to the estimation of structural parameters rather than values of endogenous variables, they yield estimates that can be regarded as raw materials, to be processed further into solutions of a wide variety of prediction problems---in particular, problems involving prediction of the effects of known changes in structure."

This passage expresses the broad thinking that econometricians have used to motivate as-if optimization.

### 4.2. Prediction under Square Loss

Haavelmo discussed application of statistical decision theory to prediction abstractly. Subsequent research using the Wald theory has focused on the case of square loss. Here the risk of a candidate predictor using sample data is the sum of the population variance of the outcome and the mean square error (MSE) of the predictor as an estimate of the mean outcome. The regret of a predictor is its MSE as an estimate of the mean. An MMR predictor minimizes maximum MSE. MMR prediction of the outcome is equivalent to minimax estimation of the population mean.

One of the earliest practical findings of statistical decision theory was reported by Hodges and Lehmann (1950). They derived the MMR predictor under square loss with data from a random sample, when the outcome has known bounded range and all sample data are observed. They assumed no knowledge of the shape of the outcome distribution. Let the outcome range be [0, 1]. Then the MMR predictor is $(\mu_N\sqrt{N} + ½)/(\sqrt{N} + 1)$, where N is sample size and $\mu_N$ is the sample average outcome.

#### 4.2.1. Prediction with Missing Data: Known Observability Rate

Extending the analysis of Hodges and Lehmann, Dominitz and Manski (2017, 2021) have studied prediction of bounded outcomes under square loss when some outcome data are missing. The former article concerns prediction of a scalar outcome. The latter studies prediction of bounded real functions of two-dimensional outcomes when data may be missing for one or both outcomes. I focus on the former case here.

It is challenging to determine the MMR predictor when data are missing. Seeking a tractable approach, we studied as-if MMR prediction. The analysis assumed knowledge of the population rate of observing outcomes, but no knowledge of the distributions of observed and missing outcomes. It used the empirical distribution of the observed data as if it were the population distribution of observable outcomes. Let K be the number of observed outcomes, which is fixed rather than random under the assumed survey design.

With no knowledge of the distribution of missing outcomes, the population mean is partially identified when the outcome is bounded. Let y be the outcome, normalized to lie in the [0, 1] interval. Let δ indicate observability of an outcome, $P(\delta = 1)$ and $P(\delta = 0)$ being the fractions of the population whose outcomes are and are not observable. Manski (1989) showed that the identification region for E(y) is the interval $[E(y|\delta = 1)P(\delta = 1), E(y|\delta = 1)P(\delta = 1) + P(\delta = 0)]$.

If this interval were known, the MMR predictor would be its midpoint $E(y|\delta = 1)P(\delta = 1) + ½P(\delta = 0)$. The interval is not known with sample data, but one can compute its sample analog and use its midpoint $E_K(y|\delta = 1)P(\delta = 1) + ½P(\delta = 0)$ as the predictor. This *midpoint predictor* is easy to compute. Dominitz and Manski (2017) showed that its maximum regret is $¼[P(\delta = 1)^2/K + P(\delta = 0)^2]$.

4.2.2. Prediction with Missing Data: Unknown Observability Rate

The above analysis assumes knowledge of the population rate of observable outcomes. A midpoint predictor remains computable when $P(\delta)$ is not known and instead is estimated by its sample analog. Derivation of an analytical expression for maximum regret appears intractable, but numerical computation is feasible. Manski and Tabord-Meehan (2017) documents an algorithm coded in STATA for numerical computation of the maximum regret of the midpoint predictor and other user-specified predictors.

The program is applicable when y is binary or is distributed continuously. In the latter case, $P_s(y|\delta = 1)$ and $P_s(y|\delta = 0)$ are approximated by Beta distributions. Subject to these restrictions on the shapes of outcome distributions, the user can specify the state space flexibly. For example, one may assume that nonresponse will be no higher than 80% or that the mean value of the outcome for nonresponders will be no lower than 0.5. One may impose no restrictions connecting the distributions $P_s(y|\delta = 1)$ and $P_s(y|\delta = 0)$, or one may bound the difference between these distributions.

Whereas Dominitz and Manski (2017) assumed a sampling process in which the number K of observed outcomes is fixed, the algorithm considers a process in which one draws at random a fixed number N of population members and sees the values of the observable outcomes. Hence, the number of observed outcomes is random rather than fixed. The midpoint predictor is $E_N(y|\delta = 1)P_N(\delta = 1) + ½P_N(\delta = 0)$. This is an unbiased and consistent estimate of the midpoint of the identification region for E(y).

Table 1 uses the program to compute the maximum regret of the midpoint predictor when y is binary. Table 2 displays maximum regret for prediction by the sample average of observed outcomes, a predictor commonly used when researchers assume that data are missing at random. Panels A and B of each table differ in their specification of the feasible outcome distributions. All distributions are feasible in Panel A. Panel B bounds the difference between the distributions of observed and missing outcomes, assuming that $-\frac{1}{2} \leq P(y = 1|\delta = 1) - P(y = 1|\delta = 0) \leq \frac{1}{2}$. Thus, prediction poses a severe identification problem in Panel A, with data on observed outcomes revealing nothing about the distribution of missing data. The problem is less severe in Panel B, where the assumption constrains the distance between the two outcome distributions. Statistical imprecision is a problem in both cases, its severity diminishing with sample size.

Each row of a table specifies a sample size, in increments of 25 from 25 to 100. Each column is a value of the observability rate, in increments of 0.1, from 0.1 to 1. Given N and $P(\delta = 1)$, the cell gives the approximate value of maximum MSE across feasible pairs of conditional outcome distributions. In each state of nature, MSE is approximated by Monte Carlo integration across 5000 simulated samples.[3] Maximum MSE across feasible outcome distributions is approximated by maximizing over a uniform grid of 100 values for each of the Bernoulli parameters $P_s(y = 1|\delta = 1)$ and $P_s(y = 1|\delta = 0)$.[4] Someone who does not know $P(\delta = 1)$ but who finds it credible to bound it can approximate maximum regret at a specified sample size by the maximum entry across the relevant column cells of the table.

To interpret the entries in the tables, keep in mind that the maximum MSE of a predictor is determined by both statistical imprecision and the identification problem created by missing data. Maximum variance decreases with sample size N. Maximum squared bias decreases with the observability rate $P(\delta = 1)$.

[3] To create a simulated sample, the program draws N observations $\{\delta_i, i = 1, \ldots, N\}$ from the distribution $P_s(\delta)$. Let $K = \sum_i \delta_i$. It draws K observations $\{y_i, i = 1, \ldots, K\}$ at random from $P_s(y|\delta = 1)$. It is possible that K = 0, so the specified predictor must be defined for this case. The program uses the N simulated realizations of δ and the K realizations of y to compute a simulated value of the predictor. For a specified positive integer T, the program repeats the above T times and uses the T simulated values of the predictor to approximate its MSE in state s.

[4] Computation of regret on a finite grid of states implies that computed maximum regret is less than or equal to true maximum regret, with equality attained if the grid contains the state yielding true maximum regret. The accuracy of the approximation may be improved by increasing grid density, at the expense of increased computation time.

Table 1A shows that identification is the dominant inferential problem when the observability rate is less than 0.7. For all sample sizes from 25 to 100, computed maximum MSE is close to $\frac{1}{4}P(\delta = 0)^2$, the value of maximum squared bias. Imprecision is a more noticeable contributor to maximum MSE when the observability rate exceeds 0.7. When the observability rate is 1, imprecision is the sole problem.

Table 1B constrains the state space to distributions satisfying $-\frac{1}{2} \leq P(y = 1|\delta = 1) - P(y = 1|\delta = 0) \leq \frac{1}{2}$. This mitigates the identification problem, yet the entries in Table 1B are essentially the same as in Table 1A. The explanation is that the states of nature generating maximum MSE when all distributions are feasible remain within the state space when the constraint is imposed. For all values of the observability rate, maximum MSE in Table 1A occurs when $P(y = 1|\delta = 1) = \frac{1}{2}$ and $P(y = 1|\delta = 0)$ equals either 0 or 1. These states are in the constrained state space of Table 1B. Hence, maximum MSE does not change.

Comparison of Tables 1A and 2A shows that, when all outcome distributions are feasible, the midpoint predictor always outperforms prediction by the sample average of observed outcomes. The maximum MSE of the former predictor is approximately ¼ the size of the latter when the observability rate is less than or equal to 0.8, and about ½ the size in the larger samples when the observability rate is 0.9. Constraining the state space in Table 2B changes the states that maximize regret for the sample-average predictor, improving its performance substantially.

## 5. Treatment Choice with Sample Data

### 5.1. Background

Econometricians and statisticians have studied treatment response in randomized trials and observational settings. Some research performs *causal inference*, without study of a decision problem. Some aims to inform treatment choice. I am concerned with the latter. The many areas of important application across the planet range from medical and other health-related treatment to educational and labor market interventions.

I use the formalization of Manski (2004). States of nature are possible distributions of treatment response for a population of observationally identical persons who are subject to treatment. The term "observationally identical" means that these persons share the same observed covariates. Groups of persons with different observed covariates are considered as separate populations.

The problem is to choose treatments for the population. Treatment response is assumed individualistic; that is, each person's outcome may depend on the treatment he receives but not on treatments received by others. Welfare is the mean outcome across the population, as in utilitarian welfare economics.

Optimal treatment is infeasible because the true distribution of treatment response is not known. Decision making uses data on the outcomes realized by a sample of the population. Some research studies data from randomized trials, and some studies observational data. Either way, statistical decision theory may be used to evaluate the performance of SDFs, called *statistical treatment rules* in this context.

A simple way to use sample data is as-if optimization. Applying the analogy principle, one acts as if the empirical outcome distribution for each treatment equals its population outcome distribution. Emulating the fact that it is optimal to chooses a treatment that maximizes the mean population outcome, one chooses a treatment that maximizes the average sample outcome. This is the empirical success (ES) rule.

When analyzing data from randomized trials, econometricians and statisticians have long used asymptotic arguments to motivate the ES rule, citing laws of large numbers and central limit theorems. A growing recent econometric literature has studied the maximum regret of the ES rule with trial data.

Consider an ideal randomized trial, where all subjects comply with assigned treatments and all realized outcomes are observed. Moreover, assume the distribution of treatment response is the same as in the population to be treated. Then the feasible states of nature are ones where, for each treatment, the population distribution of counterfactual outcomes equals that of realized outcomes.

Manski (2004) used a large-deviations inequality for sample averages of bounded outcomes to derive an upper bound on the maximum regret of the ES rule in problems of choice between two treatments. Generalizing to problems with multiple treatments, Manski and Tetenov (2016) used large deviations inequalities to bound the maximum regret of the ES rule. Stoye (2009) showed that in trials with moderate sample size, the ES rule either exactly or approximately minimizes maximum regret in cases with two treatments and a balanced design. Hirano and Porter (2009, 2019) showed that the ES rule is asymptotically optimal in a formal decision-theoretic sense. Manski and Tetenov (2016, 2019, 2021) developed algorithms for numerical computation of the maximum regret of the ES rule when outcomes are binary. Kitagawa and Tetenov (2018) studied a generalization of the ES rule, called *empirical welfare maximization*, intended for application when the set of feasible treatment policies is constrained in various ways. Also see the early related work on empirical risk minimization (e.g., Vapnik, 1999).

Econometricians have long analyzed data on realized treatments and outcomes in settings where treatments are chosen purposefully rather than randomly. Haavelmo (1944) put it this way (p. 7): "the economist is usually a rather passive observer with respect to important economic phenomena; he usually does not control the actual collection of economic statistics. He is not in a position to enforce the prescriptions of his own designs of ideal experiments."

An important contribution of early econometrics was to recognize that, when treatments are chosen purposefully, distributions of counterfactual and realized outcomes need not coincide.[5] Haavelmo (1943) showed this in the context of a linear model with homogeneous treatment response and the subsequent literature has generalized the finding substantially. Wanting to avoid the assumption of random treatment selection, econometricians have studied many models that use other assumptions to point-identify distributions of treatment response.

Recent applications of Wald's statistical decision theory to treatment choice have mainly studied decision making with data from ideal trials. Exceptions are Manski (2007), Stoye (2012), and Athey and Wager (2019), each of which studies maximum regret for rules that choose between two treatments. Manski (2007) studied trials with selective attrition and treatment choice with observational data. Stoye (2012) examined trials with some forms of imperfect internal or external validity. Athey and Wager (2019) studied choice with observational data when the set of feasible treatment policies is constrained in various ways. Whereas Manski (2007) and Stoye (2012) provided finite-sample analysis in settings where average treatment effects are partially identified, Athey and Wager performed asymptotic analysis under assumptions that give point identification.

[5] The term "analysis of treatment response" has become widespread since the 1990s, but it was not used in early writing on econometrics. A central focus was identification and estimation of models of jointly determined treatments and outcomes. The mathematical notation typically defined symbols only for realized treatments and outcomes, leaving implicit the idea of potential outcomes under counterfactual treatments. See Manski (1995, Chapter 6) and Angrist, Graddy, and Imbens (2000) for discussions that connect the early and recent literatures.

This paper presents new analysis of the maximum regret of rules for treatment with observational data when average treatment effects are partially identified. Section 5.2 develops algebraic findings for the limit case where one knows the distribution of realized outcomes in the study population. Section 5.3 reports numerical findings for treatment with sample data.

## 5.2. Treatment Choice with Knowledge of the Distribution of Realized Outcomes

I consider an observational study with two treatments {a, b} and outcomes taking values in [0, 1]. Each member of the study population has potential outcomes [y(a), y(b)]. Binary indicators [δ(a), δ(b)] denote whether these outcomes are observable. Realized outcomes are observable, but counterfactual ones are not, so the possible indicator values are [δ(a) = 1, δ(b) = 0] and [δ(a) = 0, δ(b) = 1]. State s denotes a possible distribution $P_s$[y(a), y(b), δ(a), δ(b)] of outcomes and observability. The problem is to choose between a and b in a treatment population with the same distribution of treatment response as the study population.

Given that realized outcomes are observed and counterfactual ones are not, P[δ(a) = 1] + P[δ(b) = 1] = 1. I use the notation p ≡ P[δ(b) = 1], with P[δ(a) = 1] = 1 – p. If 0 < p < 1, observation asymptotically reveals the true values of P[y(a)|δ(a) = 1], P[y(b)|δ(b) = 1], and p. Observation is uninformative about the counterfactual distributions P[y(a)|δ(a) = 0] and P[y(b)|δ(b) = 0]. This limit setting has been studied in partial identification analysis of treatment response, as in Manski (1990). I proceed likewise here.

5.2.1. MMR Treatment Choice

Manski (2007), Proposition 1 proved a simple result that holds when the planner can make a fractional treatment allocation, assigning fraction $z \in [0, 1]$ of the treatment population to treatment b and $1 - z$ to treatment a. Let all distributions of realized and counterfactual outcome distributions be feasible. Let $P[y(a)|\delta(a) = 1]$, $P[y(b)|\delta(b) = 1]$, and p be known. Then the unique MMR allocation is

$$(10) \qquad z_{MMR} = E[y(b)|\delta(b) = 1]\cdot p + \{1 - E[y(a)|\delta(a) = 1]\}(1 - p).$$

Its maximum regret is $z_{MMR}(1 - z_{MMR})$.

It is often the case that a planner cannot make a fractional allocation, being constrained by laws or by norms calling for "equal treatment of equals" to provide the same treatment to all members of the population; see Manski (2009). Moreover, a planner may be constrained to use a deterministic rather than randomized treatment rule. Thus, the planner may be required to make a binary choice between $z = 0$ and $z = 1$, Analogously, the planner chooses between treatment actions $c = a$ and $c = b$. The new analysis performed here studies this setting.

Welfare in state s in this binary choice problem equals $\max\{E_s[y(b)], E_s[y(a)]\}$ or $\min\{E_s[y(b)], E_s[y(a)]\}$. Section 2.3 showed that, in states where treatment a is better, the regret of an SDF is the product of the sampling probability that the rule commits a Type I error and the loss in welfare when choosing b. Similarly, in states where b is better, regret is the probability of a Type II error times the loss in welfare when choosing a. Thus, regret in state s is $R_{cs}\cdot|E_s[y(b)] - E_s[y(a)]|$. The treatment rule being singleton and deterministic, the error probability can only equal 0 or 1. Thus, regret is either 0 or $|E_s[y(b)] - E_s[y(a)]|$.

With outcomes having range [0, 1], $E[y(a)|\delta(a) = 0]$ and $E[y(b)|\delta(b) = 0]$ can take any values in this interval. Hence, by the Law of Iterated Expectations, the feasible values of $E[y(a)]$ and $E[y(b)]$ are

$$(11a) \quad E[y(a)] \in \big[E[y(a)|\delta(a) = 1]\cdot(1 - p),\ E[y(a)|\delta(a) = 1]\cdot(1 - p) + p\big],$$

(11b) $E[y(b)] \in \left[E[y(b)|\delta(b) = 1]\cdot p,\ E[y(b)|\delta(b) = 1]\cdot p + (1 - p)\right]$.

With choice of treatment a, regret is zero when $E_s[y(b)] < E_s[y(a)]$ and positive when $E_s[y(b)] > E_s[y(a)]$. Maximum regret occurs when $E_s[y(a)|\delta(a) = 0] = 0$ and $E_s[y(b)|\delta(b) = 0] = 1$. Then regret is

(12a) $R_{as}\cdot|E_s[y(b)] - E_s[y(a)]| = E[y(b)|\delta(b) = 1]\cdot p + \{(1 - E[y(a)|\delta(a) = 1]\}\cdot(1 - p) = z_{MMR}$.

With choice of b, maximum regret is

(12b) $R_{bs}\cdot|E_s[y(b)] - E_s[y(a)]| = E[y(a)|\delta(a) = 1]\cdot(1 - p) + \{1 - E[y(b)|\delta(b) = 1]\}\cdot p = 1 - z_{MMR}$.

Thus, treatment b minimizes maximum regret if $z_{MMR} \geq ½$ and treatment a if $z_{MMR} \leq ½$.

This finding quantifies how constraining the planner to make a deterministic singleton treatment choice rather than a fractional allocation reduces welfare. The MMR value subject to the constraint is $\min(z_{MMR}, 1 - z_{MMR})$. When fractional allocations are permitted, it is $z_{MMR}(1 - z_{MMR})$.

### 5.2.2. Maximum Regret of the ES Rule

A common assumption made when interpreting observational data posits that counterfactual and realized outcome distributions coincide, as in an ideal trial. Then treatment choice with the ES rule is optimal when the distribution of realized outcomes is known. However, the ES rule need not perform as well with larger state spaces. I examine the setting in which all outcome distributions are feasible.

Given knowledge of the distribution of realized outcomes, the ES rule chooses treatment b if $E[y(b)|\delta(b) = 1] > E[y(a)|\delta(a) = 1]$ and a if $E[y(b)|\delta(b) = 1] < E[y(a)|\delta(a) = 1]$. Either choice is consistent

with the rule if $E[y(b)|\delta(b) = 1] = E[y(a)|\delta(a) = 1]$.[6]

When $p = ½$, the ES rule coincides with the deterministic singleton rule that minimizes maximum regret. We found above that treatment b minimizes maximum regret if $z_{MMR} \geq ½$ and treatment a if $z_{MMR} \leq ½$. When $p = ½$, expression (10) reduces to $z_{MMR} = ½\{E[y(b)|\delta(b) = 1] - E[y(a)|\delta(a) = 1]\} + ½$.

When $p \neq ½$, the ES and MMR rules yield the same treatment choice in some cases but different choices in others. Both treatments are consistent with the ES rule if $E[y(b)|\delta(b) = 1] = E[y(a)|\delta(a) = 1]$. Both are consistent with the MMR rule when $z_{MMR} = ½$. I focus on cases where neither equality holds.

When $E[y(a)|\delta(a) = 1] > E[y(b)|\delta(b) = 1]$, the ES rule chooses treatment a. By (12a), maximum regret is $z_{MMR}$. When $E[y(b)|\delta(b) = 1] > E[y(a)|\delta(a) = 1]$, the rule chooses b. By (12b), maximum regret is $1-z_{MMR}$.

5.2.3. Illustration: Sentencing and Recidivism

To illustrate, I use the Manski and Nagin (1998) analysis of sentencing and recidivism of juvenile offenders in the state of Utah. The feasible treatments are sentencing options. Judges in Utah have had the discretion to order varying sentences for juvenile offenders. Some offenders have been sentenced to residential confinement (treatment a) and others have been given sentences with no confinement (treatment b). A possible policy would be to replace judicial discretion with a mandate that all offenders be confined. Another would be to mandate that no offenders be confined.

To compare these mandates, we took the outcome of interest to be recidivism in the two-year period following sentencing. Let $y = 1$ if an offender commits no new offense and $y = 0$ otherwise. No new offense was interpreted as treatment success, and commission of a new offense was interpreted as failure.

We obtained data on the sentences received and the recidivism outcomes realized by all male offenders in Utah born from 1970 through 1974 and convicted of offenses before they reached age 16. The data reveal that 11 percent of the offenders were sentenced to confinement and that 23 percent of these persons did not offend again in the two years following sentencing. The remaining 89 percent were sentenced to non-confinement and 41 percent of these persons did not offend again. Thus, P[y(a)|δ(a) = 1] = 0.23, P[y(b)|δ(b) = 1] = 0.41, and p = 0.89.

[6] Maximum regret of the ES rule when $E[y(b)|\delta(b) = 1] = E[y(a)|\delta(a) = 1]$ differs depending on whether a planner in this setting chooses a specified treatment or randomizes. Randomizing yields smaller maximum regret than choosing a specified treatment.

Reviewing the criminology literature on sentencing and recidivism, we found little research on sentencing practices and disparate predictions of treatment response. Hence, we performed partial identification analysis of treatment response, assuming no knowledge of counterfactual outcomes. This makes the present analysis applicable.

Equation (12a) shows that the maximum regret of treatment a is (0.41)(0.89) + 0.11 – (0.23)(0.11) = 0.45. Equation (12b) shows that the maximum regret of b is (0.23)(0.11) + 0.89 − (0.41)(0.89) = 0.55. Hence, treatment a minimizes maximum regret.

Suppose one assumes that Utah judges have sentenced offenders randomly to treatments and b. One then might use the ES rule to choose between the two. Given that 0.41 > 0.23, the result is choice of treatment b. Thus, the ES rule selects the treatment that is inferior from the minimax-regret perspective.

5.3. Treatment Choice with Observational Sample Data

The above analysis assumes knowledge of p, E[y(a)|δ(a) = 1], and E[y(b)|δ(b) = 1]. Suppose that one only observes realized treatments and outcomes in a random sample of the population. Sample data are informative, but they do not reveal population distributions. Hence, the state space has the higher-dimensional form $\{P_s[y(a)|\delta(a) = 1], P_s[y(a)|\delta(a) = 0], P_s[y(b)|\delta(b) = 1], P_s[y(b)|\delta(b) = 0], p_s, s \in S\}$.

Considering a planner who can make a fractional treatment allocation, Manski (2007), Proposition 2 proved that choosing the sample analog of $z_{MMR}$ as the treatment allocation yields the same maximum regret as does $z_{MMR}$. Here is the reasoning. The sample analog of $z_{MMR}$ is

$$(13) \qquad z_N = E_N[y(b)|\delta(b) = 1]\cdot p_N + \{1 - E_N[y(a)|\delta(a) = 1]\}(1 - p_N)$$
$$= E_N[y(b)\cdot\delta(b)] - E_N[y(a)\cdot\delta(a)] + (1 - p_N).$$

The second equality shows that $E(z_N) = z_{MMR}$. This and the fact that welfare is linear in $z_{MMR}$ imply that the finite-sample maximum regret achieved by $z_N$ equals the maximum regret achieved by $z_{MMR}$.

This finding extends to settings where a planner is not permitted to make a fractional allocation but can make a randomized singleton treatment choice. Consideration of randomized rules is inherently necessary when contemplating treatment choice with sample data because the data are themselves randomly drawn. It is easy to modify $z_N$ to obtain an MMR randomized singleton rule. Let u be a uniform random variable drawn independently of the sample data. Consider the rule $z_{Nu} \equiv 1[u \leq z_N]$. Then $E(z_{Nu}|z_N) = z_N$ and $E(z_{Nu}) = E[E(z_{Nu}|z_N)] = E(z_N) = z_{MMR}$.

Now suppose that a planner is constrained to make treatment choice a deterministic singleton function of the empirical distribution of realized outcomes and treatments. Thus, randomization with an independent uniform random variable is not permitted. The ES rule remains feasible. Another possibility, which I call the asymptotic minimax-regret (AMMR) rule, chooses treatment b if $z_N > ½$ and a if $z_N < ½$, with a specified default choice made if $z_N = ½$.

Algebraic computation of the maximum regret of the AMMR and ES rules appears intractable, a difficulty being that state-dependent error probabilities generally do not have explicit forms. Litvin and Manski (2021) documents an algorithm coded in STATA for numerical computation of the maximum regret of these and other user-specified treatment rules in settings with binary outcomes. Coding for the AMMR, ES, and some decision rules using instrumental variables is built in, including default treatment choices when a rule does not yield a unique choice.

The program uses Monte Carlo integration to approximate state-dependent error probabilities for a specified rule. Maximum regret is approximated by computing regret on a grid that discretizes the state space. The state space specifies feasible values for the five Bernoulli distributions $\{P_s[y(a)|\delta(a) = 1], P_s[y(a)|\delta(a) = 0], P_s[y(b)|\delta(b) = 1], P_s[y(b)|\delta(b) = 0], P_s[\delta(b) = 1)], s \in S\}$. A user can place a flexible set of constraints on the feasible distributions. One may place lower and/or upper bounds on the values of the Bernoulli probabilities. One may impose no restrictions connecting distributions of realized and counterfactual outcome, or one may bound the difference between these distributions.

Tables 3 and 4 use the program to compute the maximum regret of the AMMR and ES rules. Panels A and B of each table differ in the feasible outcome distributions. All distributions are feasible in Panel A. Panel B bounds the difference between distributions of realized and counterfactual outcomes, assuming that $-\frac{1}{2} \leq P[y(t) = 1|\delta(t) = 1] - P[y(t) = 1|\delta(t) = 0] \leq \frac{1}{2}$, $t \in \{a, b\}$.

Each row of a table specifies a sample size N, in increments of 25 from 25 to 100. Each column is a value of p, in increments of 0.1 from 0.5 to 9. It is unnecessary to consider values of p below 0.5 because the state space in each panel views the two treatments symmetrically; hence, maximum regret is the same for p and 1 – p.

Given values for N and p, a cell entry presents the approximate value of maximum regret across feasible values of $\{P[y(t)|\delta(t) = 1], P[y(t)|\delta(t) = 0]\}$, $t \in \{a, b\}$. In each state of nature, regret is approximated by Monte Carlo integration across 5000 simulated samples.[7] Maximum regret across feasible Bernoulli distributions is approximated by maximizing over a grid of 25 values for each Bernoulli parameter. Someone who does not know p but who finds it credible to bound it can approximate maximum regret at a specified sample size by the maximum entry across the relevant column cells of the table.

Table 3A has many interesting features. Observe that the maximum regret of the AMMR rule does not vary with p. Moreover, for each value of p, maximum regret rises rather than falls with sample size N, increasing from about 0.34 at N = 25 to 0.40 at N = 100. Part of the explanation for both features is that sampling variation makes the rule more randomized for small N and less so for large N. In the polar case N = 1, the AMMR rule coincides with the randomized singleton rule $z_N$, which has maximum regret ¼ for all values of p.[8] As $N \rightarrow \infty$, the AMMR rule approaches the MMR deterministic singleton rule, which has maximum regret ½ for all values of p.[9]

[7] To create a simulated sample, the program draws N observations $\{\delta_i, i = 1, \ldots, N\}$ from distribution $P_s[\delta(b) = 1]$. Let $K = \sum_i \delta_i$. It draws K observations $\{y_i, i = 1, \ldots, K\}$ from $P_s[y(b)|\delta(b) = 1]$ and N – K from $P_s[y(a)|\delta(a) = 1]$. It is possible that K = 0 or N, so the specified treatment rules must be defined for these cases. The program uses the simulated realizations of $(\delta, y)$ to compute a simulated treatment choice. For a specified positive integer T, the program repeats the above T times and uses the T simulated treatment choices to approximate regret in state s.

Table 3B constrains the state space to bound the difference between realized and counterfactual outcomes distributions. This mitigates the identification problem. The restriction reduces maximum regret moderately when p = 0.5, but only negligibly when p = 0.9.

Tables 4A and 4B consider the ES rule. Section 5.2.2 considered this rule when the distribution of realized outcomes is known. It was shown that, depending on this distribution, maximum regret equals $z_{MMR}$ or $1 - z_{MMR}$. Holding p fixed, maximum regret across all values of E[y(a)|δ(a) = 1] and E[y(b)|δ(b) = 1] is max (p, 1 – p). Table 4A shows that this is also maximum regret for the finite-sample version of the ES rule.[10] Table 4B shows that restricting the state space to bound the difference between realized and counterfactual distributions reduces maximum regret. The findings are somewhat subtle due to the small-sample effect of constraining the state space, with different behavior for p = 0.5 and 0.6 than for $p \geq 0.7.6$.

## 6. Conclusion

To reiterate the central theme of this paper, use of statistical decision theory to evaluate econometric models is conceptually coherent and simple. A planner specifies a state space listing all the states of nature deemed feasible. One evaluates the performance of an SDF by the state-dependent vector of expected welfare that it yields. Decisions using models are evaluated in this manner. One evaluates model-based SDFs by their performance across the state space, not across the model space.

[8] When N = 1, $z_N$ can only take the value 0 or 1. Hence, $z_N = 1[z_N > ½]$. It was found above that, for any value of N, the maximum regret of $z_N$ given knowledge of the distribution of realized outcomes is $z_{MMR}(1 - z_{MMR})$. Maximum regret across all {p, E[y(a)|δ(a) = 1], E[y(b)|δ(b) = 1]} is ¼, which occurs when $z_{MMR}$ = ½. This occurs for every p when E[y(a)|δ(a) = 1] = E[y(b)|δ(b) = 1] = ½.

[9] It was found above that, given knowledge of the distribution of realized outcomes, the maximum regret of the MMR deterministic singleton rule is min ($z_{MMR}$, $1 - z_{MMR}$). Maximum regret across all {p, E[y(a)|δ(a) = 1], E[y(b)|δ(b) = 1]} is ½, which occurs when $z_{MMR}$ = ½. This occurs for every p when E[y(a)|δ(a) = 1] = E[y(b)|δ(b) = 1] = ½.

[10] The ES rule does not benefit in small samples from the randomizing effect of random sampling because the states yielding maximum regret are polar ones where E[y(a)|δ(a) = 1] and E[y(b)|δ(b) = 1] equal 0 or 1. Holding p fixed, the supremum of regret across {E[y(a)|δ(a) = 1], E[y(b)|δ(b) = 1]} such that E[y(a)|δ(a) = 1] > E[y(b)|δ(b) = 1] is p when p > ½ and 1 – p when p < ½. The former occurs when E[y(a)|δ(a) = 1] = 1 and E[y(b)|δ(b) = 1] → 1 and the latter when E[y(b)|δ(b) = 1] = 0 and E[y(a)|δ(a) = 1] → 0. The supremum of regret across {E[y(a)|δ(a) = 1], E[y(b)|δ(b) = 1]} such that E[y(b)|δ(b) = 1] > E[y(a)|δ(a) = 1] is again p when p > ½ and 1 – p when p < ½. The former occurs when E[y(a)|δ(a) = 1] = 0 and E[y(b)|δ(b) = 1] → 0 and the latter when E[y(b)|δ(b) = 1] = 1 and E[y(a)|δ(a) = 1] → 1.

The primary challenge to use of statistical decision theory is computational. Recall that, in his discussion sketching application of statistical decision theory to prediction, Haavelmo (1944) remarked that such application (p. 111): "although simple in principle, will in general involve considerable mathematical problems and heavy algebra." Many mathematical operations that were infeasible in 1944 are tractable now, as a result of advances in analytical and numerical methods. The advances include important theorems yielding useful large deviations bounds and local-asymptotic approximations, as well as revolutionary developments in computer capabilities that enable approximate numerical solution of decision problems. Hence, it has increasingly become possible to use statistical decision theory when performing econometric research that aims to inform decision making. Future advances should continue to expand the scope of applications.

## Appendix: Further Discussion of Statistical Decision Theory

### A.1. Bayes Decisions

Considering contexts where one wants to minimize loss rather than maximize welfare, research in statistical decision theory often refers to criterion (4) as minimization of *Bayes risk*. This term may seem odd given the absence of any reference in (4) to Bayesian inference. Criterion (4) simply places a subjective distribution π on the state space ex ante and optimizes the resulting subjective average welfare. No posterior distribution computed after observation of data appears in the criterion.

Justification for use of the word *Bayes* when considering (4) rests on a mathematical result relating this criterion to conditional Bayes decision making. The conditional Bayes approach calls on one to first perform Bayesian inference, which uses the likelihood function for the observed data to transform the prior distribution on the state space into a posterior distribution, without reference to a decision problem. One then chooses an action that maximizes posterior subjective average welfare.

See, for example, the seminal text of DeGroot (1970) on *Optimal Statistical Decisions* or discussions of applications to randomized trials in articles such as Spiegelhalter, Freedman, and Parmar (1994), Cheng, Su, and Berry (2003), and Scott (2010). The conditional Bayes perspective has long been used to study not only static decisions but also sequential decisions, which are formalized as dynamic programming problems. These include problems of sequential experimentation, sometimes called *bandit problems*.

As described above, conditional Bayes decision making is unconnected to Wald's frequentist statistical decision theory. However, suppose that the set of feasible statistical decision functions is unconstrained and that certain regularity conditions hold. Then it follows from Fubini's Theorem that the conditional Bayes decision for each possible data realization solves Wald's problem of maximization of subjective average welfare. See Berger (1985, Section 4.4.1) for general analysis and Chamberlain (2007) for application to a linear econometric model with instrumental variables. On the other hand, Kitagawa and Tetenov (2018) and Athey and Wager (2019) study treatment-choice problems in which the set of feasible decision functions is constrained. Wald's criterion (4) need not yield the same actions as conditional Bayes decision making in these constrained settings.

The equivalence of Wald's criterion (4) and conditional Bayes decisions is a mathematical result that holds under specified conditions. Philosophical advocates of the conditional Bayes paradigm go beyond the mathematics. They assert as a self-evident axiom that decision making should condition on observed data and should not perform frequentist thought experiments that contemplate how statistical decision functions perform in repeated sampling; see, for example, Berger (1985, Chapter 1).

Considering the mathematical equivalence of minimization of Bayes risk and conditional Bayes decisions, Berger asserted that that the conditional Bayes perspective is normatively "correct" and that the Wald frequentist perspective is "bizarre." He stated (p. 160):

> "Note that, from the conditional perspective together with the utility development of loss, the *correct* way to view the situation is that of minimizing $\rho(\pi(\theta|x), a)$. One should condition on what is known, namely x . . . . and average the utility over what is unknown, namely $\theta$. The desire to minimize $r(\pi, \delta)$ would be deemed rather bizarre from this perspective."

In this passage, a is an action, x is data, $\theta$ is a state of nature, $\pi(\theta|x)$ is the posterior distribution on the state space, $\rho$ is posterior loss with choice of action a, $\delta$ is a statistical decision function, $\pi$ is the prior distribution on the state space, and $r(\pi, \delta)$ is the Bayes risk of $\delta$.

I view Berger's normative statement to be overly enthusiastic for two reasons. First, the statement does not address how decisions should be made when part of the decision is choice of a procedure for collection of data, as in experimental or sample design. Such decisions must be made ex ante, before collecting the data. Hence, frequentist consideration of the performance of decision functions across possible realizations of the data is inevitable. Berger recognized this in his chapter on "Preposterior and Sequential Analysis."

Second, the Bayesian prescription for conditioning decision making on sample data presumes that the planner feels able to place a credible subjective prior distribution on the state space. However, Bayesians have long struggled to provide guidance on specification of priors and the matter continues to be controversial. See, for example, the spectrum of views regarding Bayesian analysis of randomized trials expressed by the authors and discussants of Spiegelhalter, Freedman, and Parmar (1994). The controversy suggests that inability to express a credible prior is common in actual decision settings.

When one finds it difficult to assert a credible subjective distribution, Bayesians may suggest use of some default distribution, called a “reference” or “conventional” or “objective” prior; see Berger (2006). However, there is no consensus on the prior that should play this role. The chosen prior affects decisions.

### A.2. Maximin and Minimax Regret

Concern with specification of priors motivated Wald (1950) to study the minimax criterion. He wrote (p. 18): “a minimax solution seems, in general, to be a reasonable solution of the decision problem when an a priori distribution . . . . does not exist or is unknown to the experimenter.”

I similarly am concerned with decision making in the absence of a subjective distribution on the state space. However, I have mainly measured the performance of SDFs by maximum regret rather than by minimum expected welfare. The maximin and MMR criteria are sometimes confused with one another, but they are equivalent only in special cases, particularly when the value of optimal welfare is invariant across states of nature. The criteria obviously differ more generally. Whereas maximin considers only the worst outcome that an action may yield across states, MMR considers the worst outcome relative to what is achievable in a given state.

Practical and conceptual reasons motivate focus on maximum regret. From a practical perspective, MMR decisions behave more reasonably than do maximin ones in the important context of treatment choice with data from randomized trials. In common settings with balanced designs, the MMR rule is well approximated by the empirical success rule, which chooses the treatment with the highest observed average outcome in the trial; see Section 5. In contrast, the maximin criterion ignores the trial data, whatever they may be. This was recognized verbally by Savage (1951), who stated that the criterion is “ultrapessimistic” and wrote (p. 63): “it can lead to the absurd conclusion in some cases that no amount of relevant experimentation should deter the actor from behaving as though he were in complete ignorance." Savage did not flesh out this statement, but it is easy to show that this occurs with trial data; see Manski (2004).

The conceptual appeal of using maximum regret to measure performance is that maximum regret quantifies how lack of knowledge of the true state of nature diminishes the quality of decisions. While the term "maximum regret" has been standard in the literature, this term is a shorthand for the maximum sub-optimality of a decision criterion across the feasible states of nature. An SDF with small maximum regret is uniformly near-optimal across all states. This is a desirable property.

Minimax regret has drawn diverse reactions from decision theorists. In a famous early critique, Chernoff (1954) observed that MMR decisions are not always consistent with the choice axiom known as independence of irrelevant alternatives (IIA). He considered this a serious deficiency, writing (p. 426):

> "A third objection which the author considers very serious is the following. In some examples, the min max regret criterion may select a strategy $d_3$ among the available strategies $d_1$, $d_2$, $d_3$, and $d_4$. On the other hand, if for some reason $d_4$ is made unavailable, the min max regret criterion will select $d_2$ among $d_1$, $d_2$, and $d_3$. The author feels that for a reasonable criterion the presence of an undesirable strategy $d_4$ should not have an influence on the choice among the remaining strategies."

This passage is the totality of Chernoff's argument. He introspected and concluded that reasonable decision criteria should always adhere to the IIA axiom, but he did not explain why he felt this way. Chernoff's view has been endorsed by some modern decision theorists, such as Binmore (2009). However, Sen (1993) argued that adherence to axioms such as IIA does not per se provide a sound basis for evaluation of decision criteria. He asserted that consideration of the context of decision making is essential.

Manski (2011) argued that adherence to the IIA axiom is not a virtue per se. What matters is how violation of the axiom affects welfare. I observed that the MMR violation of the IIA axiom does not yield choice of a dominated SDF. The MMR decision is always undominated when it is unique. There generically exists an undominated MMR decision when the criterion has multiple solutions. Hence, I concluded that violation of the IIA axiom is not a sound rationale to dismiss minimax regret.

### A.3. Choosing the Welfare Function, State Space, and Decision Criterion

In principle, statistical decision theory views the welfare function, state space, and decision criterion as separate meta-choices made the planner. The welfare function expresses what the planner wants to achieve. The state space lists all states that he believes might occur. The decision criterion expresses how he wants to cope with uncertainty. Statistical decision theory views these meta-choices as predetermined rather than as matters to be studied within the theory.

Arguably, planners might choose welfare functions, state spaces, and decision criteria jointly rather separately. A practical reason is that some joint choices yield decision problems that are easier to solve than others. Hence, planner might find it desirable to make a joint choice that yields tractable solutions.

A fundamental issue for meta-choice is that some choices may not be implementable. Consider, for example, maximization of subjective expected welfare when the state space is nonparametric. It is mathematically delicate to specify subjective distributions on nonparametric spaces. For example, the familiar notion of a flat prior may not be well-defined.

Another issue is that implementable choices which initially seem appealing may yield solutions that seem "unattractive" in some sense. An example is Savage's conclusion, discussed above, that maximin treatment choice with data from a randomized trial is unattractive because it ignores the data entirely.

Stoye (2009) studies a more subtle case of decisions that ignore trial data entirely. This occurs when members of the population have an observable covariate that is continuously distributed across persons and when the state space does not constrain how treatment response varies with the covariate. His analysis focuses on MMR, but his method of proof uses the equivalence of MMR to maximization of subjective expected welfare with a certain worst-case prior. Hence, his result also holds for this case of maximization of subjective expected welfare.

Table 1A: Maximum MSE of Midpoint Predictor
Unrestricted Outcome Distributions

| N | P(δ = 1) | | | | | | | | | |
|---|---|---|---|---|---|---|---|---|---|---|
| | 0.1 | 0.2 | 0.3 | 0.4 | 0.5 | 0.6 | 0.7 | 0.8 | 0.9 | 1 |
| 25 | 0.1963 | 0.1567 | 0.1219 | 0.0917 | 0.0658 | 0.0451 | 0.0292 | 0.0184 | 0.0120 | 0.0105 |
| 50 | 0.1956 | 0.1554 | 0.1200 | 0.0891 | 0.0632 | 0.0422 | 0.0258 | 0.0140 | 0.0072 | 0.0052 |
| 75 | 0.1953 | 0.1552 | 0.1196 | 0.0887 | 0.0624 | 0.0409 | 0.0243 | 0.0125 | 0.0055 | 0.0034 |
| 100 | 0.1952 | 0.1548 | 0.1191 | 0.0881 | 0.0620 | 0.0404 | 0.0237 | 0.0119 | 0.0048 | 0.0025 |

Table 1B: Maximum MSE of Midpoint Predictor
$-½ \leq P(y = 1|\delta = 1) - P(y = 1|\delta = 0) \leq ½$

| N | P(δ = 1) | | | | | | | | | |
|---|---|---|---|---|---|---|---|---|---|---|
| | 0.1 | 0.2 | 0.3 | 0.4 | 0.5 | 0.6 | 0.7 | 0.8 | 0.9 | 1 |
| 25 | 0.1962 | 0.1567 | 0.1214 | 0.0912 | 0.0658 | 0.0451 | 0.0292 | 0.0184 | 0.0120 | 0.0105 |
| 50 | 0.1956 | 0.1553 | 0.1199 | 0.0890 | 0.0631 | 0.0422 | 0.0258 | 0.0140 | 0.0072 | 0.0052 |
| 75 | 0.1952 | 0.1552 | 0.1192 | 0.0885 | 0.0622 | 0.0408 | 0.0242 | 0.0125 | 0.0055 | 0.0034 |
| 100 | 0.1951 | 0.1548 | 0.1189 | 0.0881 | 0.0620 | 0.0402 | 0.0237 | 0.0119 | 0.0048 | 0.0025 |

Table 2A: Maximum MSE of Prediction with Average Observed Outcome
Unrestricted Outcome Distributions

| N | P(δ = 1) | | | | | | | | | |
|---|---|---|---|---|---|---|---|---|---|---|
| | 0.1 | 0.2 | 0.3 | 0.4 | 0.5 | 0.6 | 0.7 | 0.8 | 0.9 | 1 |
| 25 | 0.7409 | 0.6150 | 0.4716 | 0.3465 | 0.2407 | 0.1543 | 0.0870 | 0.0389 | 0.0148 | 0.0105 |
| 50 | 0.7782 | 0.6166 | 0.4714 | 0.3464 | 0.2408 | 0.1542 | 0.0869 | 0.0387 | 0.0102 | 0.0052 |
| 75 | 0.7793 | 0.6159 | 0.4713 | 0.3463 | 0.2406 | 0.1541 | 0.0868 | 0.0387 | 0.0098 | 0.0034 |
| 100 | 0.7779 | 0.6151 | 0.4710 | 0.3462 | 0.2404 | 0.1541 | 0.0868 | 0.0387 | 0.0098 | 0.0025 |

Table 2B: Maximum MSE of Prediction with Average Observed Outcomes
$-½ \leq P(y = 1|\delta = 1) - P(y = 1|\delta = 0) \leq ½$

| N | P(δ = 1) | | | | | | | | | |
|---|---|---|---|---|---|---|---|---|---|---|
| | 0.1 | 0.2 | 0.3 | 0.4 | 0.5 | 0.6 | 0.7 | 0.8 | 0.9 | 1 |
| 25 | 0.3189 | 0.2230 | 0.1599 | 0.1172 | 0.0836 | 0.0576 | 0.0375 | 0.0231 | 0.0143 | 0.0105 |
| 50 | 0.2654 | 0.1875 | 0.1411 | 0.1033 | 0.0732 | 0.0486 | 0.0298 | 0.0165 | 0.0082 | 0.0052 |
| 75 | 0.2406 | 0.1773 | 0.1345 | 0.0987 | 0.0693 | 0.0457 | 0.0273 | 0.0141 | 0.0063 | 0.0034 |
| 100 | 0.2278 | 0.1719 | 0.1303 | 0.0957 | 0.0673 | 0.0442 | 0.0261 | 0.0132 | 0.0054 | 0.0025 |

Table 3A: Maximum Regret of the AMMR Rule
Unrestricted Outcome Distributions

| N | p | | | | |
|---|---|---|---|---|---|
| | 0.5 | 0.6 | 0.7 | 0.8 | 0.9 |
| 25 | 0.3416 | 0.3441 | 0.3421 | 0.3428 | 0.3469 |
| 50 | 0.3747 | 0.3782 | 0.3773 | 0.3792 | 0.3777 |
| 75 | 0.3903 | 0.3887 | 0.3899 | 0.3892 | 0.3899 |
| 100 | 0.4023 | 0.4021 | 0.4011 | 0.4026 | 0.4022 |

Table 3B: Maximum Regret of the AMMR Rule
$-\frac{1}{2} \leq P[y(t) = 1|\delta(t) = 1] - P[y(t) = 1|\delta(t) = 0] \leq \frac{1}{2}, t \in \{a, b\}$

| N | p | | | | |
|---|---|---|---|---|---|
| | 0.5 | 0.6 | 0.7 | 0.8 | 0.9 |
| 25 | 0.2772 | 0.2905 | 0.3041 | 0.3178 | 0.3313 |
| 50 | 0.3126 | 0.3290 | 0.3421 | 0.3534 | 0.3646 |
| 75 | 0.3277 | 0.3397 | 0.3523 | 0.3648 | 0.3774 |
| 100 | 0.3402 | 0.3577 | 0.3713 | 0.3852 | 0.3927 |

Table 4A: Maximum Regret of the ES Rule
Unrestricted Outcome Distributions

| N | p | | | | |
|---|---|---|---|---|---|
| | 0.5 | 0.6 | 0.7 | 0.8 | 0.9 |
| 25 | 0.5000 | 0.6000 | 0.7000 | 0.8000 | 0.9000 |
| 50 | 0.5000 | 0.6000 | 0.7000 | 0.8000 | 0.9000 |
| 75 | 0.5000 | 0.6000 | 0.7000 | 0.8000 | 0.9000 |
| 100 | 0.5000 | 0.6000 | 0.7000 | 0.8000 | 0.9000 |

Table 4B: Maximum Regret of the ES Rule
$-\frac{1}{2} \leq P[y(t) = 1|\delta(t) = 1] - P[y(t) = 1|\delta(t) = 0] \leq \frac{1}{2}, t \in \{a, b\}$

| N | p | | | | |
|---|---|---|---|---|---|
| | 0.5 | 0.6 | 0.7 | 0.8 | 0.9 |
| 25 | 0.2791 | 0.3000 | 0.3500 | 0.4000 | 0.4500 |
| 50 | 0.3086 | 0.3075 | 0.3500 | 0.4000 | 0.4500 |
| 75 | 0.3280 | 0.3243 | 0.3500 | 0.4000 | 0.4500 |
| 100 | 0.3381 | 0.3375 | 0.3500 | 0.4000 | 0.4500 |